\documentclass[twocolumn,prl,amsmath,amssymb]{revtex4}
\usepackage{graphicx}
\usepackage{booktabs}
\usepackage{multirow}
\usepackage{amssymb}
\usepackage[version=3]{mhchem} 
\usepackage{graphicx}
\usepackage{amsmath}
\usepackage{indentfirst}
\usepackage{pstricks}
\usepackage{verbatim}
\usepackage{bm}
\usepackage{color}   

\begin{document}

\title{Electroluminescence in Single Layer MoS$_{2}$}
\author{R. S. Sundaram$^{1}$, M. Engel$^{2}$, A. Lombardo$^{1}$, R. Krupke$^{2,3}$, A. C. Ferrari$^{1}$, Ph. Avouris$^{4,*}$, M. Steiner$^{4,*}$,}
\affiliation{$^1$Department of Engineering, University of Cambridge, Cambridge CB3 0FA, UK\\
$^2$Institute of Nanotechnology, Karlsruhe Institute of Technology, 76021 Karlsruhe, Germany\\
$^3$Institut f\"{u}r Materialwissenschaft, Technische Universit\"{a}t Darmstadt, 64287 Darmstadt, Germany\\
$^4$IBM Thomas J. Watson Research Center, Yorktown Heights, New York 10598, USA\\
$^*$Corresponding author: avouris@us.ibm.com, msteine@us.ibm.com}

\begin{abstract} We detect electroluminescence in single layer molybdenum disulphide (MoS$_{2}$) field-effect transistors built on transparent glass substrates. By comparing absorption, photoluminescence, and electroluminescence of the same MoS$_{2}$ layer, we find that they all involve the same excited state at 1.8eV. The electroluminescence has pronounced threshold behavior and is localized at the contacts. The results show that single layer MoS$_{2}$, a direct band gap semiconductor, is promising for novel optoelectronic devices, such as 2-dimensional light detectors and emitters.
\end{abstract}

\maketitle
Molybdenum disulphide (MoS$_{2}$), a layered quasi-2 dimensional (2d) chalcogenide material\cite{frindt_jap_1966}, is subject of intense research because of its electronic\cite{kis_nn_2011} and optical properties\cite{heinz_prl_2010}, such as strong photoluminescence (PL)\cite{heinz_prl_2010,fenwang_nl_2010}, controllable valley and spin polarization\cite{heinz_nn_2012} and a large on-off ratio field effect transistors (FETs)\cite{kis_nn_2011}. A single layer of MoS$_{2}$ (1L-MoS$_{2}$) consists of two planes of hexagonally arranged S atoms linked to a hexagonal plane of Mo atoms via covalent bonds\cite{heinz_prl_2010,pollmann_prb_2001,wold_prb1987,groot_prb_1987,mattheiss_prb_1973}. In the bulk, individual MoS$_2$ layers are held together by weak van der Waals forces \cite{pollmann_prb_2001,wold_prb1987,groot_prb_1987,mattheiss_prb_1973}. This property has been exploited in lubrication technology\cite{lieber_apl_1991} and, more recently, has lead to the isolation of 1L-MoS$_{2}$\cite{kis_nn_2011,heinz_prl_2010,fenwang_nl_2010,novoselov_s_2004}. While bulk MoS$_{2}$ is a semiconductor with an indirect band gap of 1.3 eV \cite{parkison_jpc_1092}, 1L-MoS$_{2}$ has a direct band gap of 1.8 eV\cite{heinz_prl_2010,fenwang_nl_2010}. The absence of interlayer coupling of electronic states at the $\Gamma$ point of the Brillouin zone in 1L-MoS$_{2}$[\onlinecite{fenwang_nl_2010,galli_jpcc_2007}] results in strong absorption and PL bands at$\sim$1.8eV (680nm)\cite{heinz_prl_2010,fenwang_nl_2010}. 1L-MoS$_{2}$ FETs have both unipolar\cite{kis_nn_2011} and ambipolar\cite{iwasa_nl_2012} transport characteristics with mobilities$>$500$cm^{2}V^{-1}s^{-1}$\cite{pdye_IEEE_2012} and on-off current ratios up to $10^{9}$[\onlinecite{pdye_acs_2012}]. This combination of electrical and optical properties suggests that 1L-MoS$_{2}$ is a promising candidate for novel optoelectronic devices, such as 2d photodetectors\cite{zhang_acs_2011,orta_oe_2012,im_nl_2012}, and light-emitting devices operating in the visible range.

Here, we report electrically excited luminescence in 1L-MoS$_{2}$ FETs, and study the underlying emission mechanism. We find that the electroluminescence occurs via hot carriers and is localized in the contacts region. The observed photoluminescence and electroluminescence arise from the same excited state at 1.8eV.
\begin{figure*}
\centerline{\includegraphics[width=170mm]{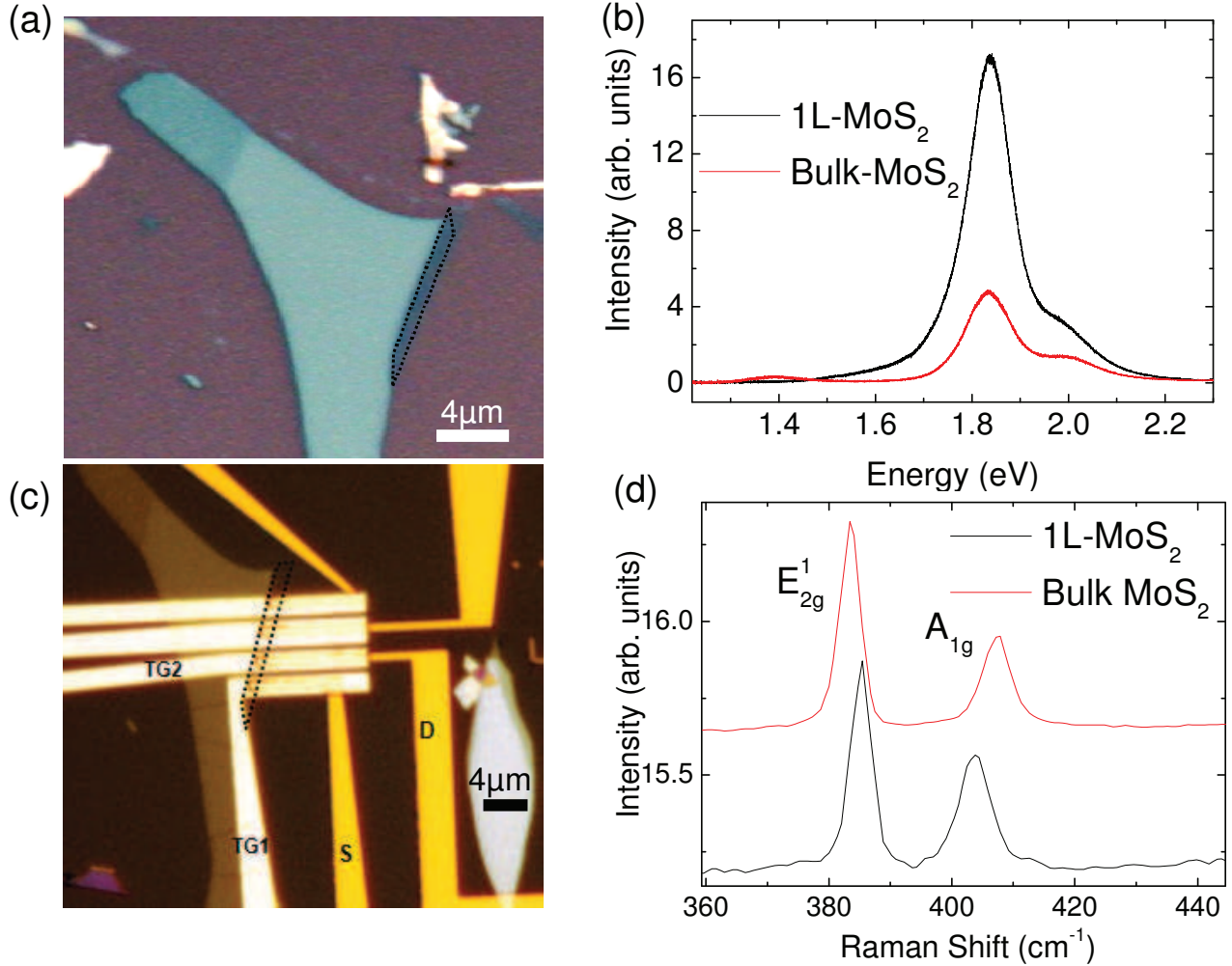}}
\caption{(a) Optical white light microscope image of a MoS$_{2}$ flake. The monolayer region is highlighted by dashed lines. (b) PL spectrum measured on (red) bulk MoS$_{2}$ and (black) 1L-MoS$_{2}$ for 514.5nm excitation. The PL is stronger for 1L-MoS$_{2}$. (c) Optical image of the MoS$_{2}$ devices with source (S), drain (D) and top gate electrodes (TG1, TG2). The 1L-MoS$_{2}$ position is highlighted by the black dashed line.(d)Raman spectra measured for 514.5nm excitation in (top) bulk MoS$_{2}$, and (bottom) 1L-MoS$_{2}$. The difference in peak positions identifies the monolayer\cite{ryu_acs_2010}}
\label{1}
\end{figure*}

1L-MoS$_{2}$ crystals are produced by micromechanical cleavage of bulk MoS$_{2}$ (Structure Probe Inc.-SPI, Natural Molybdenite) on 100nm SiO$_{2}$. As for the case of graphene\cite{ferrari_nl_2007}, interference allows visibility and counting the number of layers, Fig.\ref{1}a. Due to the different dielectric properties, an optimum thickness of 100nm SiO$_{2}$ is well suited for MoS$_{2}$\cite{kis_n_2007}. The presence of monolayers is then confirmed by performing PL measurements, Fig.\ref{1}b. The PL spectrum of 1L-MoS$_{2}$ exhibits two bands at 2eV and 1.8eV (Fig.\ref{1}b) associated with excitonic transitions at the K point of the Brillouin zone\cite{fenwang_nl_2010}. The energy difference of 0.2eV has been attributed to the degeneracy breaking of the valence band due to spin-orbit coupling \cite{fenwang_nl_2010,wold_prb1987,groot_prb_1987,schuller_apl_2011}. As compared to bulk MoS$_{2}$, Fig.\ref{1}b, 1L-MoS$_{2}$ does not have a peak at 1.4eV\cite{heinz_prl_2010,fenwang_nl_2010}, associated with the indirect band gap\cite{parkison_jpc_1092}. In addition 1L-MoS$_{2}$ exhibits a stronger PL intensity compared to bulk MoS$_{2}$\cite{heinz_prl_2010,fenwang_nl_2010} due to the direct band gap. Another evidence for 1L-MoS$_{2}$ comes from the analysis of the Raman spectrum, Fig.1d. The peak at$\sim$385cm$^{-1}$ corresponds to the in plane (E$^{1}_{2g}$) mode\cite{ryu_acs_2010}, while that at $\sim$404 cm$^{-1}$ is attributed to the out of plane (A$_{1g}$) mode\cite{ryu_acs_2010}. The E$^{1}_{2g}$ mode softens and A$_{1g}$ mode stiffens with increasing layer thickness\cite{ryu_acs_2010}, similar to what happens for other layered materials, where the bond distance changes with number of layers\cite{arenal_nl_2006}. The frequency difference between these two modes can be used as a signature of 1L-MoS$_{2}$\cite{ryu_acs_2010}.

1L-MoS$_{2}$ flakes are then transferred onto glass substrates by using a poly(methyl methacrylate) (PMMA) based transfer technique, similar to that previously used to transfer graphene onto optical fibre cores\cite{ferrari_np_2010}. This process involves spin coating two layers of 495K PMMA and one layer of 950K PMMA on the substrate where flakes are deposited. The samples are subsequently immersed in de-ionized (DI) water at 90$^{\circ}$C for 1h, resulting in the detachment of the polymer film, due to the intercalation of water at the polymer-SiO$_{2}$ interface. MoS$_{2}$ flakes stick to the PMMA, and can thus be removed from the original substrate and mechanically transferred onto glass substrates\cite{ferrari_np_2010}. In order to manufacture a device with split top gates, we use e-beam lithography to define source and drain contacts, followed by thermal evaporation of Au (50nm) with a Cr adhesion layer (2nm). This Cr thickness is sufficient for the adhesion of 50nm Au pads, typically used as source-drain contacts in MoS$_{2}$ transistors\cite{kis_nn_2011}. The gate dielectric is then made via atomic layer deposition (ALD) of Al$_{2}$O$_{3}$ (30nm). This thickness was previously used for electrostatic p-n junctions in nanotubes, and offers a compromise between film uniformity and gate capacitance\cite{engel_acs_2012}. Top gate electrodes are also made by evaporating 50nm Au with a 2nm Cr (Fig.\ref{1}c).
\begin{figure}
\centerline{\includegraphics[width=80mm]{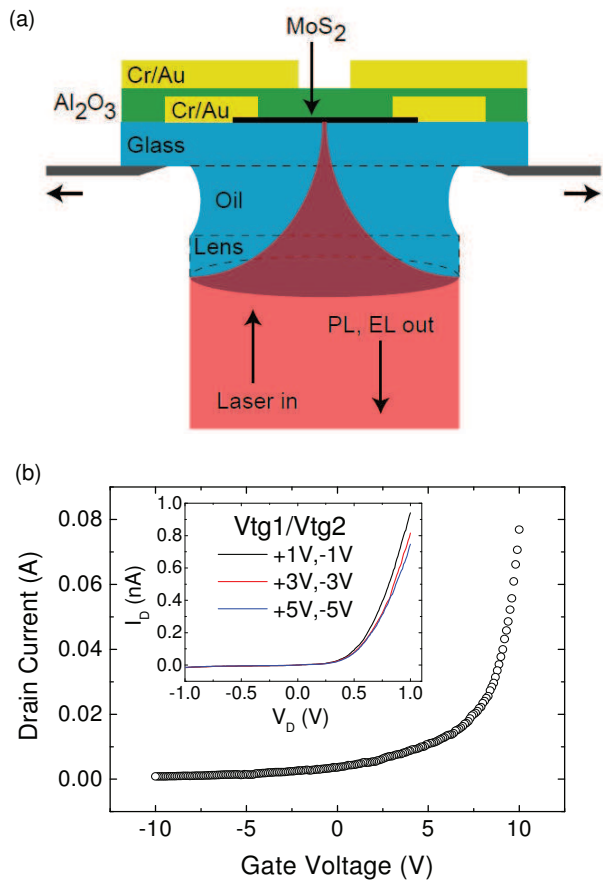}}
\caption{(a)Schematic of a top-gated MoS$_{2}$ FET and the optical setup. (b) Electrical transfer characteristics of a 1L-MoS$_{2}$ FET at bias voltage V$_D$=200mV. (Inset) Electrical output characteristics of a 1L-MoS$_{2}$ FET. The gate voltage pairs (V$_{TG1}$/V$_{TG2}$) applied in the V$_D$-sweeps are also indicated}
\label{2}
\end{figure}

Photocurrent data are acquired by converting the short-circuit photocurrent between the source and drain (ground) electrodes into a voltage signal by using a current preamplifier and a source meter synchronized with a controlling computer and an optical scanning system, using the setup sketched in Fig.2a. This combines an electrical transport measurement system with an inverted optical microscope equipped with a multi-axis stage for raster scanning the devices with respect to a tightly focused laser beam (diameter 250nm, $\lambda$=514.5nm, immersion objective, NA=1.25)\cite{engel_acs_2012,avouris_nl_2011}. Optical spectroscopy is done via a liquid-nitrogen cooled, back-illuminated, deep-depleted charge coupled device (CCD) and a 300 grooves/mm grating, as for Refs.\onlinecite{engel_acs_2012,avouris_nl_2011,engel_nc_2012}.
\begin{figure}
\centerline{\includegraphics[width=80mm]{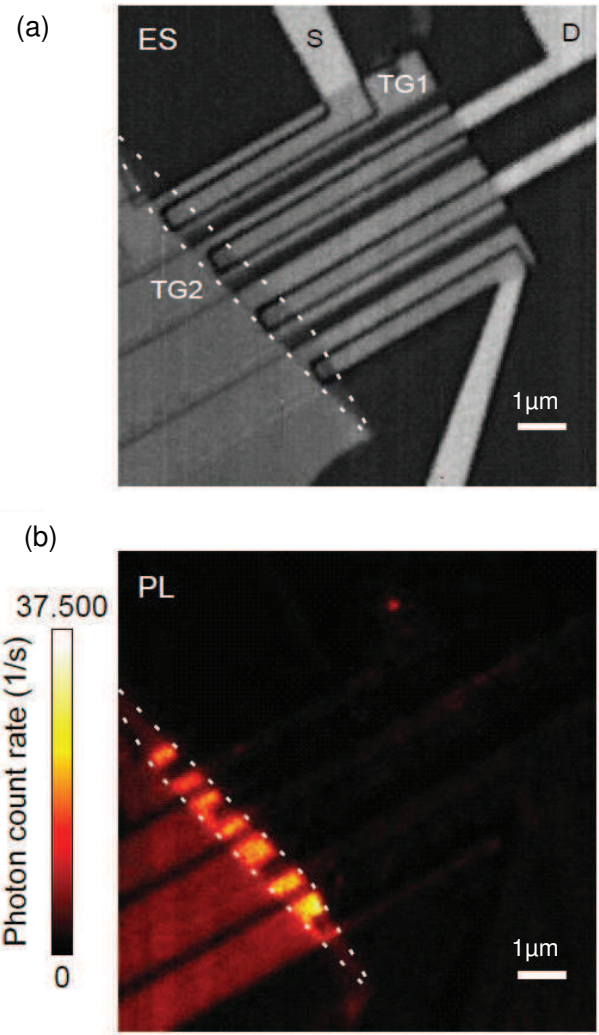}}
\caption{(a) The elastic scattering (ES) image taken from the underside of the MoS$_{2}$ device array reveals the position of source (S) and drain (D) contacts and the top gate electrodes (TG1, TG2). (b) The PL image of the same area reveals the 1L-MoS$_{2}$ position within the device array highlighted in (a) by white dashed lines.}
\label{3}
\end{figure}

Fig.\ref{3}a plots an elastic scattering image of the device, taken by raster scanning a 20x20$\mu$$m^{2}$ area with a 50nm step size, and acquired in confocal mode with laser illumination at 514.5nm, through the immersion objective, and with a single photon counting module in the detection path. PL images are recorded in confocal mode (laser power density P$_{Laser}<$100kW/cm$^{2}$) through a band pass filter spectrally centered at 700nm. Fig.\ref{3}b plots the PL map of the same device imaged in Fig.3a. The PL intensity is higher at the 1L-MoS$_{2}$ position, as expected\cite{heinz_prl_2010,fenwang_nl_2010}. The PL intensity appears further enhanced underneath the metal gates, a result of the higher collection efficiency due to reflection. In contrast, the PL is quenched at the MoS$_{2}$-Cr/Au interface, with no significant PL at the source and drain contacts.

Fig.\ref{2}b plots the electrical transfer characteristics acquired by sweeping both top gates simultaneously at a drain-source voltage V$_{D}$=200mV. The device has a 2M$\Omega$ on-state resistance and on-off current ratio $10^{3}$. The inset in Fig.\ref{2}b shows the behavior of the output current for different settings of split gate voltages, having either equal or opposite polarity. While the efficacy of the split gate in current modulation is limited, we have some non-linearity in the I-V curves, with a reproducible dependence on varying split gate voltages. In the following, we use photocurrent microscopy to study the gate dependence of the electrostatic potential in the device.
\begin{figure}
\centerline{\includegraphics[width=80mm]{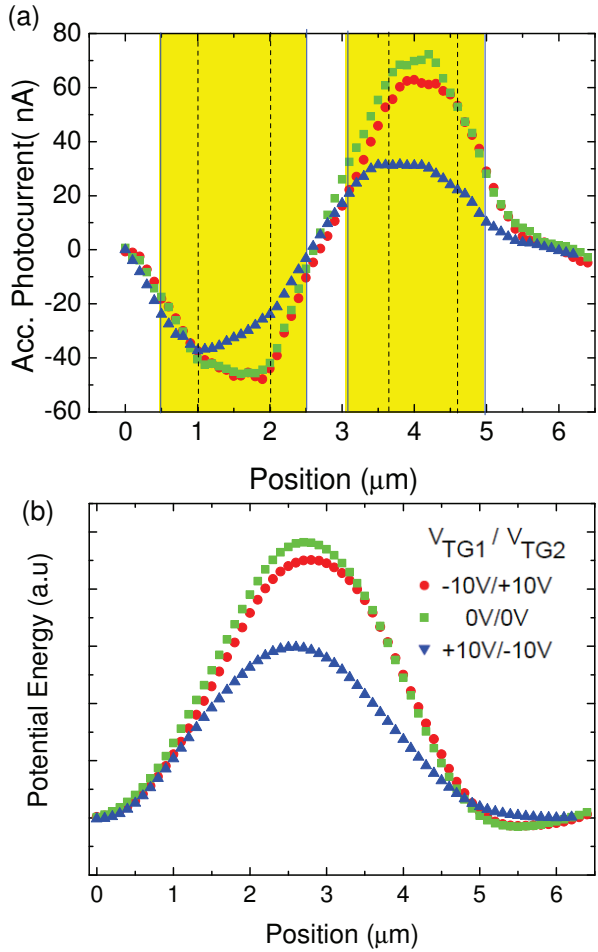}}
\caption{Normalized photocurrent measured from the underside of a MoS$_{2}$ device for three representative split gate voltage pairs (V$_{TG1}$/V$_{TG2}$): (-10V/+10V) red circles, (0V/0V) green squares, (+10V/-10V) blue triangles. The curves represent the total photocurrent perpendicular to the channel direction, based on the measured photocurrent images and a linear background correction. The position of source and drain contacts is indicated by dashed black lines, the position of the top gates by the golden areas. (b) Internal electrostatic potential cross sections of the MoS$_{2}$ device for three split gate voltage pairs. The curves are obtained by numerically integrating the experimental photocurrent data in (a).}
\label{4}
\end{figure}

Exfoliated 1L-MoS$_{2}$ behaves as an n-doped semiconductor, with a Fermi level at 4.7eV($\phi_{MoS_{2}}$)\cite{kis_nn_2011,zhang_acs_2011,geim_pnas_2005}. The intrinsic doping has been assigned to halogen (Cl or Br) impurities in natural MoS$_{2}$ crystals\cite{zhang_acs_2011}. If MoS$_{2}$ is brought in contact with Cr/Au, having work functions $\phi_{Cr/Au}$= 4.8eV\cite{riviere_1969} to 5.1eV \cite{zhang_acs_2011}, a Schottky barrier ($\phi_{SB}$) is formed with a height of 100 to 400meV [$\phi_{SB}$= $\phi_{Cr/Au}-\phi_{MoS_{2}}$] and we expect a strong photocurrent response at the contacts, similar to other 1d and 2d nanostructures, such as carbon nanotubes\cite{kern_sm_2007}, silicon nanowires\cite{lauhon_nl_2009} and graphene\cite{mews_acs_2012}. Fig.\ref{4}a plots the accumulated photocurrent cross sections recorded by raster scanning the device with respect to the focused laser beam. After image acquisition, the photocurrent is measured along the direction perpendicular to the device channel, and overlaid with the position of drain-source contacts and top gates for three representative top gate voltage settings. The photocurrent response at the contact edges is mainly due to the Schottky barriers at the MoS$_{2}$-Cr/Au interface. Ideally, we expect that p- and n-type regions in the MoS$_{2}$ channel could be created by using electrostatic doping through the application of appropriate split gate voltages, as previously shown in nanotubes\cite{engel_acs_2012,kinoshita_oe_2010,mueller_nn_2010} and graphene\cite{gabor_s_2011}. However, the gate efficiency in our devices is too low, and even for the highest electrostatic potential gradient (+10V/-10V) we observe only a weak effect on the measured photocurrent amplitude (Fig.\ref{4}a). While we find that the Schottky barrier for electron injection can be modulated through the application of the respective split gate voltages, we do not observe significant hole currents in both photocurrent and electrical transport measurements. As a result, our gating configuration is inefficient for creating a p-n junction within the device, as evidenced in the electrostatic potential profiles in Fig.\ref{4}b, generated by numerically integrating the accumulated photocurrent amplitudes in Fig.4a. This has important implications for the electrical detection and generation of light emission within the present device configuration. The conversion efficiencies of photons to carriers (or carriers to photons) cannot be controlled by the gates in the present case and are largely determined by the internal device electrostatics.

Since electrons and holes cannot be injected independently into the MoS$_{2}$ channel, we exploit hot carrier processes for measuring the electroluminescence (EL) spectrum of 1L-MoS$_{2}$. At high bias, electrons injected into the conduction band should experience a strong band bending at the MoS$_{2}$-metal contact, with generation of excitons via impact excitation, a process extensively studied in semiconducting nanotubes\cite{avouris_nph_2008}. Additionally, we expect 1L-MoS$_{2}$ to heat up significantly at high bias, similar to graphene\cite{engel_nc_2012,freitag_nl_2009,bericaud_prl_2010}, which could result in a thermal population of the emitting state.
\begin{figure}
\centerline{\includegraphics[width=80mm]{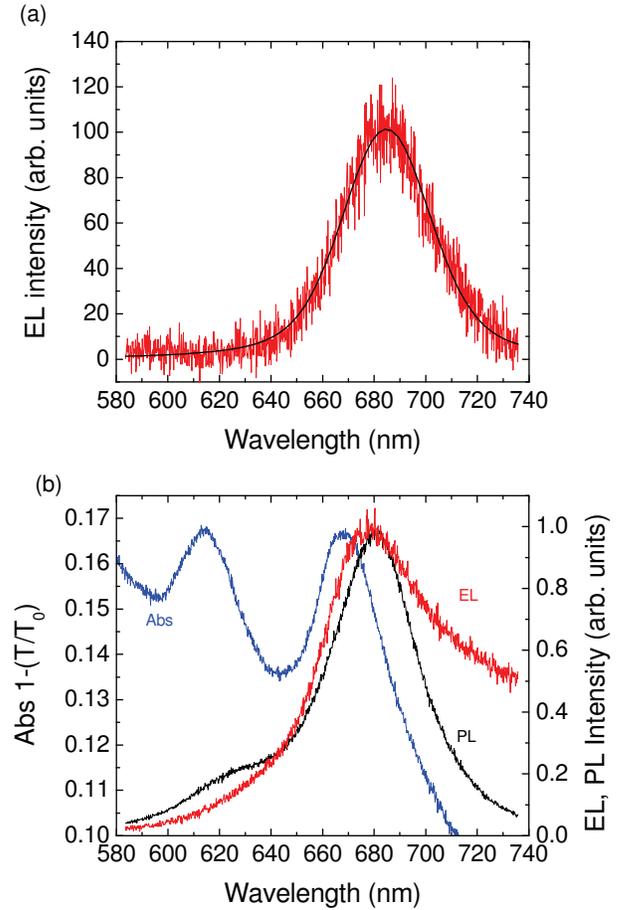}}
\caption{(a) EL of a 1L-MoS$_{2}$ device measured at V$_{D}$=5V and I$_{D}$=100$\mu$A with a Voigt fit. (b) Absorption (Abs), EL, and PL spectra on the same 1L-MoS$_{2}$. The EL spectrum is measured at V$_{D}$=8V and I$_{D}$=164$\mu$A}
\label{5}
\end{figure}

In our experiment, we determine the optimum EL bias by tuning the source drain voltage while mapping the EL emission by means of a single photon counting detector. Fig.\ref{5}a shows the EL spectrum of 1L-MoS$_{2}$. The spectral distribution has a full-width-at-half-maximum (FWHM)$\sim$40nm, and a peak position$\sim$685nm. For comparison, we plot a high-bias EL spectrum along with PL and absorption spectra in Fig.5b. The two main features at 610 and 670nm are associated with the A and B excitons of MoS$_{2}$\cite{heinz_prl_2010,fenwang_nl_2010,heinz_nn_2012}. Their positions correspond well with the observed PL peaks at 620 and 680nm. A Stokes shift$\sim$10nm separates the positions of absorption and PL peaks. We assign this to surface interaction of 1L-MoS$_{2}$ within the inhomogeneous dielectric environment (substrate SiO$_{2}$, gate dielectric Al$_{2}$O$_{3}$), which in turn influences the exciton binding energy due to screening of the electron-hole Coulomb interaction\cite{korn_pss_2012}. Importantly, the peak position in the EL spectrum matches the PL peak at$\sim$680nm, evidencing that EL and PL emission involve the same excited state, i.e. the B exciton. However, the PL feature at 620nm cannot be observed in the EL spectrum, highlighting that the excitation mechanisms in PL and EL are different. While high electrical bias causes spectral broadening and increased thermal background in the EL spectrum (see Fig.\ref{5}b) we do not observe significant spectral shifts.
\begin{figure}
\centerline{\includegraphics[width=80mm]{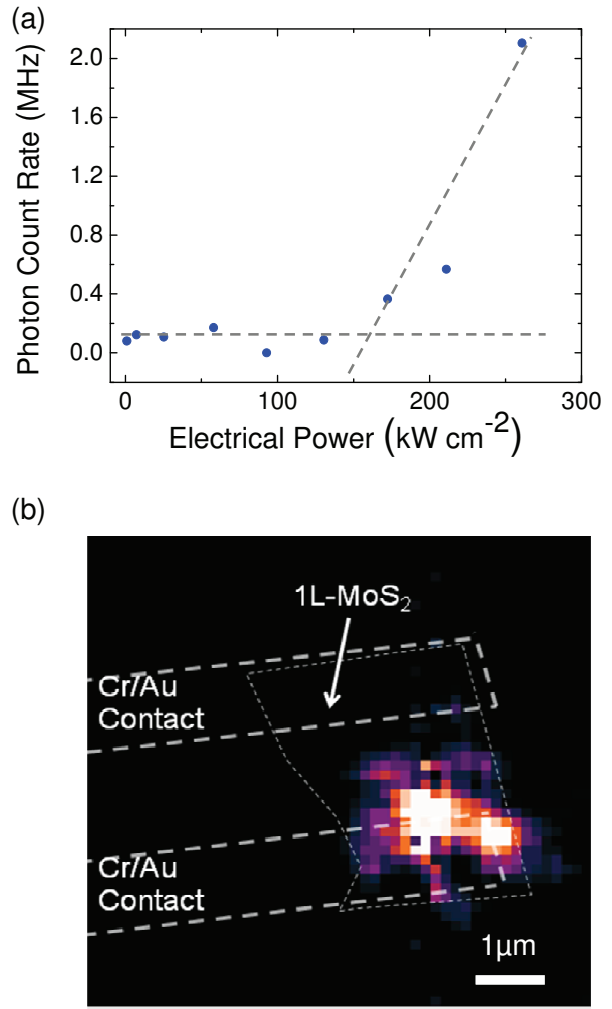}}
\caption{(a) EL count rate as function of injected electrical power density for a different  1L-MoS$_{2}$ device. Each data point represents the sum of all photon counts recorded within an image of the entire device area at a given electrical power density. The dark count rate is~0.9MHz. The dashed lines are a guide to the eyes. (b) ) False color image showing EL emission in the vicinity of a contact edge. The positions of Cr/Au contacts are highlighted by thick dashed lines (white) and the MoS2 layer is indicated by thin dashed lines (grey). The device is biased at V$_{D}$=4V and I$_{D}$=1.8mA.}
\label{6}
\end{figure}

In order to investigate the threshold behavior and the efficiency of EL generation we plot in Fig.\ref{6}a the integrated light intensity as function of electrical power density $P_{EL}=\frac{V_DI_D}{L_CW_C}$ injected into the MoS$_{2}$ channel (channel length L$_{C}$=1.5$\mu$m, channel width W$_{C}$= 2.3$\mu$m). We observe significant light emission only above a threshold power density of 150kW/cm$^{2}$. The reason is that the electrons need to acquire sufficient kinetic or thermal energy for the generation of excitons. The EL threshold bias hence depends on the exciton binding energy and the thermal properties of the channel material, as well as the specifics of the semiconductor-metal contacts of the actual device. The exciton-to-phonon conversion efficiency is calculated by dividing the integrated photon count rate (Fig.\ref{6}a) by the quantum efficiency of the detector, the fraction of light detected within the solid angle (based on the objective NA) and the losses of light when it passes through the objective and mirrors along the optical path. This gives us the total photon flux originating from the sample. When putting this in relation to the current, i.e. carriers per unit time, we arrive at a conservative estimate of the conversion efficiency of$\sim$10$^{-5}$. For a comparison, this is at least an order of magnitude lower than thus far reported for individual semiconducting single walled nanotubes\cite{mueller_nn_2010}. The conversion efficiency could be enhanced significantly by creating a p-n structure that enables threshold-less carrier recombination.

Since hot carrier effects rely on band deformation\cite{freitag_nl_2004}, the efficiency of exciton generation through impact excitation and thermal population should be maximized at the positions where the carrier injection occurs, i.e. at the source and drain electrodes. We hence expect EL not homogeneously radiated from the 1L-MoS$_{2}$, but spatially localized near the contacts. In order to map the EL spatial distribution within the 1L-MoS$_{2}$ device, we raster scan EL images of the biased device with a single photon counting module through a band pass filter centered at 700nm. From an elastic laser scattering image acquired with the same detector, we are able to locate the position of source and drain contacts with high precision. By overlaying the contact positions in the EL image shown in Fig.\ref{6}b, we find that the EL emission is indeed localized at one of the metal contacts.

In order to exploit 1L-MoS$_{2}$ in practical optoelectronic devices, the efficiency of light detection and emission needs to be significantly enhanced. Novel device designs are needed to improve yield, and control charge carrier injection and extraction, such as the use of highly efficient gates to create electrostatic p-n junctions. An alternative could be the use of strong doping via polymer electrolytes\cite{sood_nn_2008} or intercalation\cite{ferrari_jacs_2007}.

In conclusion, single layer MoS$_{2}$ transistors can detect and emit visible light. Both photoluminescence and electroluminescence arise from the same excited state at 1.8eV. Better electrostatic gating techniques are needed to improve control and efficiency of light emission and detection in optoelectronic devices made of MoS$_{2}$.

We acknowledge funding from EU grants NANOPOTS, GENIUS, EPSRC grants EP/GO30480/1 and EP/G042357/1, a Royal Society Wolfson Research Merit Award, and St. Edmund's College, Cambridge.

\end{document}